# In situ photoemission electron spectroscopy study of nitrogen ion implanted AISI-H13 steel

L.F. Zagonel, C.A. Figueroa, F. Alvarez

Instituto de Física "Gleb Wataghin", Universidade Estadual de Campinas, Unicamp, 13083-970, Campinas, São Paulo, Brazil

## Abstract

In this paper we report the effect of hydrogen on the structural properties of AISI-H13 steel nitrogen-implanted samples in low oxygen partial pressure atmosphere. The samples were implanted in a high vacuum chamber by using a broad ion beam source. The $H_2^+/N_2^+$ ion composition of the beam was varied and the surface composition studied in situ by photoemission electron spectroscopy (XPS). The samples were also ex situ analyzed by X-ray diffraction and scanning electron microscopy (SEM), including energy-dispersive spectroscopy measurements. It was found that hydrogen has the effect of modifying the amount of retained nitrogen at the surfaces. This result shows that hydrogen plays a role beyond the well-established effect of oxygen etching in industrial machines where vacuum is relatively less well controlled. Finally, an optimum concentration of 20–40% $[H_2]/[H_2+N_2]$ ion beam composition was determined to obtain maximum nitrogen incorporation on the metal surface.

## 1. Introduction

Plasma nitriding is a well-established technique broadly applied to improve surface properties of steels [1]. Nevertheless, some important physical mechanisms involved in the process are still not well understood. Indeed, the microscopic mechanisms in nitriding processes and their effects on the metal structure continue to be an important subject of study [2–4]. The use of hydrogen in plasma nitriding is a standard procedure to increase nitrogen penetration in steel [5]. The presence of stable oxides on the surface blocks nitrogen diffusion and the beneficial effect of hydrogen attributed to etching effects on the oxides compounds. However, the role of hydrogen is sometimes suggested to go beyond the chemical etching effects. Moreover, some authors claim that the presence of hydrogen has effects on the plasma structure, influencing nitrogen incorporation in the bulk of the metal [6,7]. Therefore, to observe specific effects of hydrogen, experiments in an oxygen-free atmosphere are mandatory.

In this paper we report a comprehensive nitriding study of AISI-H13 steel in accurately controlled conditions. The samples were implanted in a high vacuum chamber (partial pressure ∼2 x$10^5$ Pa) by using a Kaufman cell fed with different nitrogen–hydrogen mixtures. The deposition chamber is attached to an ultrahigh vacuum chamber for in situ photoemission electron spectroscopy analysis (XPS) for composition and structural analysis. Also, the structure evolution of the material was studied by ex situ X-ray diffraction. Morphology studies, by ex situ SEM analysis, show the nitrogen diffusion via grain boundaries. Experimental results show that 20–40% hydrogen partial pressure increases nitrogen concentration in the metal bulk. This experimental finding confirms that hydrogen plays a role beyond the well-established oxygen etching.

## 2. Experimental

The samples studied in the present work are rectangular (20x15x2 mm) slices from the same AISI-H13 steel lot. Table 1 displays the material composition as determined by XPS analysis in sputtered cleaned raw samples. The nitriding process was performed on a high vacuum system with a base pressure always better than $2 \times 10^5$ Pa ($P[O_2] \sim 10^6$ Pa) using a Kaufman cell. This system is attached to a UHV chamber where in situ X-ray photo-electron spectroscopy (XPS) measurements are performed. More details of the implanting system can be found elsewhere [8]. To carry out this study, all system parameters were maintained constant and the hydrogen and nitrogen flux proportion accurately adjusted. These gases are admitted into the Kaufman cell through independent mass flow controllers. The hydrogen partial pressure, $C_H=[H_2]/[N_2+H_2]$, was varied from 0 to 80 at.% maintaining fixed the total chamber pressure ($\sim 1.4 \; 10^2$ Pa). Other important parameters maintained constant during ion implantation are: ion beam energy, 600 eV; ion beam effective current density: 2 mA/cm$^2$ (measured with a faraday cup); temperature: 450 8C; implantation time: 2 hours. At this energy, both nitrogen and hydrogen will be dissociated and shallow implanted [2,9,10]. Indeed, SRIM simulations show that at 300 eV, nitrogen and hydrogen will reach 1 and 5 nm depths, respectively [11]. These characteristic depths show that a thin oxide layer may indeed be an efficient barrier for nitrogen incorporation and diffusion. Moreover, hydrogen passivation on this first 5 nm layer might increase both, nitrogen retention (up to saturation concentrations) and mobility (through alloying elements traps screening) [12]. Finally, after the nitriding process, the samples are transferred to the ultrahigh vacuum chamber ($\sim 5 \times 10^7$ Pa) for XPS measurements.

The neutral species contained in the beam are difficult to measure. However, we can make an attempt to estimate their concentrations. Conservation laws forbid neutralization of the ions in the gas phase at the working pressures used in the present study [13]. Therefore, the origin of the neutral species in the beam could stem from two sources: (1) neutralization due to electrons injected by the filament neutralizer of the Kaufman cell, and (2) secondary electrons emitted due to ion impact on grounded parts of the Kaufman cell. In the experimental conditions reported in this paper the beam neutralizer was not used. Therefore, this contribution to the formation of neutral species is not present. Regarding secondary electron emission, it can be neglected. Indeed, the secondary emission coefficient of 100 eV $N_2^+$ ions hitting molybdenum is c~0.032 [14,15]. Considering that c weakly depends on energy, one can estimate that at most ~3.2% of the $N_2^+$ ion will be neutralized in the ion beam. The XPS spectra were obtained by using the 1486.6 eV photons from an Al target (Kα-line) and a VG-CLAMP-2 electron analyzer. The total apparatus resolution was ~0.85 eV (line-width plus analyzer). The relative atomic composition at the sample surfaces was determined by integrating the core level peaks, properly weighted by the photo-emission cross-section. As is well known, XPS gives information of the outmost atomic material layers (~0.5 nm) [16]. Crystallographic information was acquired with the usual h–2h mode X-ray diffraction (XRD) with the Cu Ka-line. Material hardness measurements were performed using a standard Vickers micro-indenter (0.05 kg load, Shimadzu) and the morphological characterization by a scanning electron microscopy (SEM) (Jeol JMS-5900LV), equipped with energy depressives spectrometry (EDS). The samples for SEM measurements were mirror-polished and 120 seconds etched in 2% nital solution. SEM images and EDS were obtained using 8 keV electron beam energy and 15 keV, respectively.

## 3. Results and discussion

### 3.1. Influence of the $H_2/[H_2+N_2]$ gaseous mixture on surface composition and hardness

Fig. 1(a) shows the nitrogen and iron concentrations obtained from XPS studies as a function of hydrogen partial pressures in the Kaufman cell in the absence of oxygen. This plot shows the nitrogen dependence retention on the surface material by the presence of $H_2^+$ in the ion beam. The curves show that a maximum (minimum) nitrogen (iron) concentration at the surfaces is obtained for 30–40% (20–30%) $H_2/[H_2+N_2]$ gaseous mixtures. This indicates that even if the plasma nitriding process takes place in a very low oxygen partial pressure chamber, some hydrogen will enhance nitrogen surface retention. The observed diminishing N concentration above around 30–40% gaseous concentration is probably due to chemical etching. Also, the increment of hydrogen partial pressure decreases the effective nitrogen ionic current, contributing to a lower nitrogen concentration on the surface. Fig. 1(b) displays surface hardness vs. gaseous mixture. Within the experimental error, hydrogen slightly reduces hardness to an almost constant value, independent of gas mixture composition. As we shall see below, however, the crystalline structure depends on the gaseous mixture. Finally, it is important to remark that the oxygen concentration measured by XPS on the surfaces was smaller than 3 at.% for all studied samples. This is expected since XPS analyses were performed in situ in a high vacuum chamber.

Table 1
AISI H13 steel composition, at.%

| Element | Fe | C | Mn | Si | Cr | Mo | V |
|---|---|---|---|---|---|---|---|
| Concentration (F0.5 at.%) | 90.5 | 0.6 | 0.3 | 0.3 | 6.7 | 1 | 0.7 |

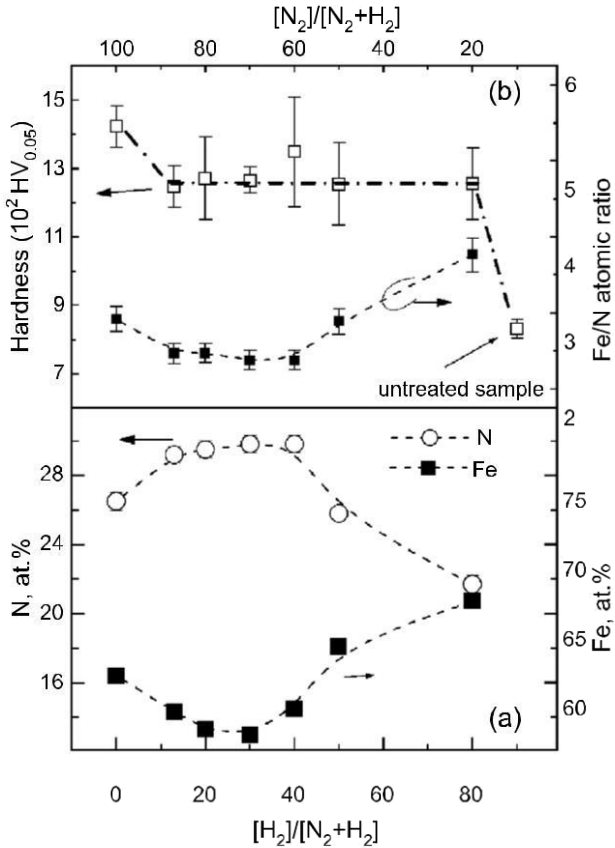

Fig. 1. (a) Nitrogen and iron surface concentration as function of gaseous nitrogen concentration feeding the ion source. (b) Frontal hardness and iron/ nitrogen ratio as function of gaseous nitrogen concentration feeding the ion source. Lines are guide to the eyes.

3.2. Influence of the $H_2/[H_2+N_2]$ gaseous mixture on the material structure

The X-ray diffraction patterns of the studied samples are enlightening about the phases present in the material. Fig. 2 shows the evolution of the crystalline material on N incorporation. A pristine sample is also included for comparison purposes. Except for sample H2-80, implanted with 80% H2/[H2+N2] gaseous mixture, the ε-phase ($Fe_{2-3}N$) is sizable in all curves [17,18]. The α-phase (bcc) is also apparent as a shoulder in some nitrated samples. The presence of these phases is consistent with the iron/nitrogen stoichiometry obtained from the XPS data after subtracting nitrogen bounded to alloying elements (see Fig. 1b). The constituent alloying elements (Cr, V, Mo) produces two effects: stabilization of the original a-phase and more retention of nitrogen in the structure [19]. In fact, depending on the nitrogen concentration, two phenomena are possible. On the one hand, at relative low N content, the α-phase is predominant and N does not exceed the solubility limit of the element in the Fe–N binary system [20]. On the other hand, at higher N content, the q-phase becomes dominant. Thus, the presence of the α-phase observed in different samples even at nitrogen concentration above the solubility limit suggests non-equilibrium nitrogen over saturated α-phase.

X-ray diffraction also reveals that samples treated with less hydrogen are more expanded (not shown) possibly explaining the greater hardness of sample H₂-00.

The photoelectron emission spectra (XPS) of the allowing elements, such as $Cr2p_{3/2}$ and $V2p_{3/2}$, show the onset of a second peak. For chromium, this second peak, associated with $Cr_{1+x}N$ (x near to 0), has a chemical shift of ~1.1 eV and is responsible for ~40% of the total signal independently of the $[H_2]/[N_2+H_2]$ mixture feeding the ion source [21]. For the sake of clarity, only the results obtained with $[H_2]/[N_2+H_2]=50\%$ are displayed in Fig. 3(a). We remark that the curve asymmetry is due to shake-up process [16]. Assuming that, after nitrogen incorporation, two bands associated to chromium environments, where used to fit the curves (see Fig. 3). The first band corresponds to the $Cr2p_{3/2}$ core electrons of nitrogen-free material and the second one to the electrons associated to $Cr2p_{3/2}$ core electrons in chromium nitrides compounds. The latest contribution is chemically shifted due to the electronegative effect of nitrogen [16].

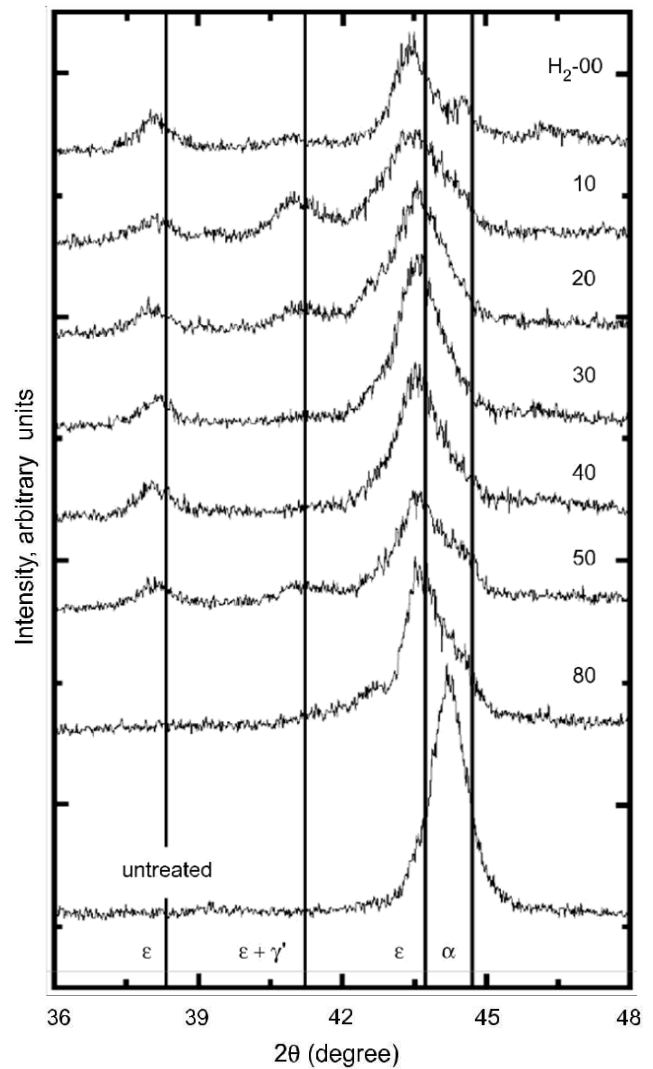

Fig. 2. X-ray difractogram of nitrated samples. The characteristic position of ε-, γ'- and α-phases are indicated.

Similar results are obtained for all studied gaseous mixtures. Also, we note that the chemical shift to higher energies observed in the photoemission electron core level binding energies is consistent with the high N electronegativity. The photoelectron emission spectrum associated with the N1s electron is a broad band centered at ~394.4 eV (Fig. 3b). This band (width ~1.9 eV) is constituted by electrons stemming from contributions of several metallic nitrides such as CrN, VN, MoN, and $Fe_{2-4}N$. We note that this band overlaps with the one associated with the spectrum of the $Mo3p_{3/2}$ electron core level. In Fig. 3, a ~10% contribution stemming from the Mo $p_{3/2}$ core electron (not shown) has been properly subtracted.

3.3. Morphology

Fig. 4 shows the SEM result (backscattered mode) obtained in sample $H_2$-50 taken from its transversal cross section. Several features are observed in this picture. First, the dark zones around grain boundaries near the surface indicate the presence of light nuclear density elements such as N. Therefore, grain boundary diffusion through the presence of defects of N is strongly suggested for the picture, explaining the relative thick diffusion layer for this sample (~30 Am). Second, the small bright spots indicate metallic nitrides precipitates. The higher density of these spots at the outmost layers of the samples is due to higher N concentrations near the sample surfaces. At the surface, therefore, the large amount of precipitates suggests a composition bulk grain change by alloying elements depletion. In order to verify this hypothesis, spatially resolved energy-dispersed X-ray spectroscopy (EDS) was performed at the borders and bulk region of sample grains. In particular, the probed zone was located at the outmost sample layers and deeper in the studied samples. At the outmost layers, the EDS results confirm the increasing (decreasing) alloying elements, such as Cr and V, at the grain boundary (bulk) as compared with normal concentration found in deeper material layers. This is understood because N extracts allowing element from the bulk grains to form the precipitates, depleting the bulk of the grains near the material surfaces. Unfortunately, SEM pictures of the studied samples do not show sizable differences in samples treated with diverse $H_2/[H_2+N_2]$ mixtures. However, the effect observed on N retention is probably due to the formation of hydrides at the grain boundary, playing an important role in the diffusion process [21–23].

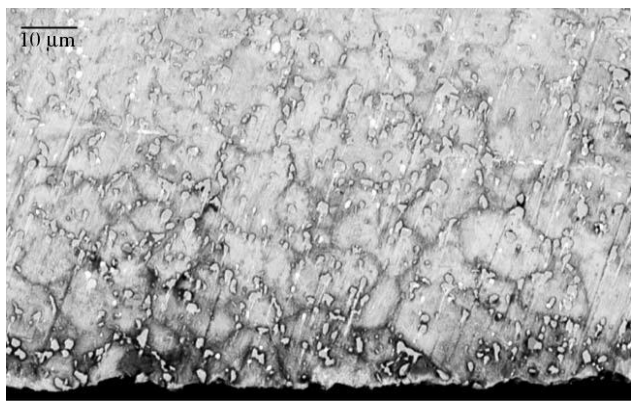

Fig. 4. SEM image in backscattered configuration (sample $H_2$-50). The dark grain contours are evidences of nitrogen presence. The bright smallest spots are metallic precipitates.

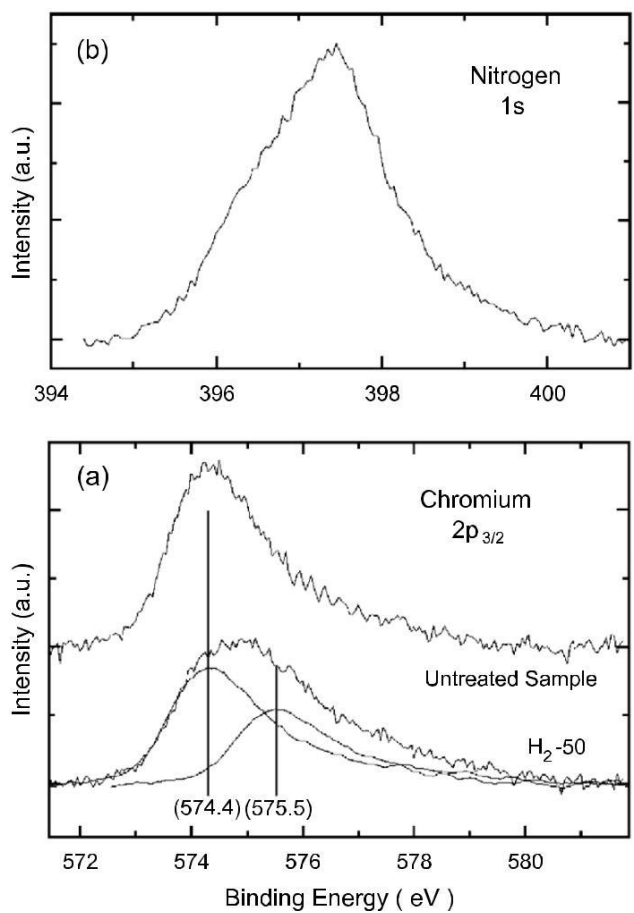

Fig. 3. (a) Spectra corresponding to the Cr2p3/2 photoelectrons for untreated and nitrided samples with 50% hydrogen feeding the ion source. It can be observed the onset of a second band, 1.1 eV chemical shifted toward increasing binding energies. (b) N 1s band formed by contribution stemming from all N chemical states.

4. Conclusion

In conclusion, in this work we have shown that, at very low oxygen partial pressure (~$10^6$ Pa), appropriated nitrogen dilution in hydrogen improves nitrogen retention at the surface of the material. This results shows that hydrogen has a specific role going beyond that of the well-established oxygen etching. Depending on the nitrogen concentration, two phenomena can occur. First, for relative low N concentration, the a-phase is stabilized. Second, at higher N surface retention, the q-phase is formed. Scanning electron microscopy confirms that N diffuses by grain boundary. Moreover, metallic nitrite precipitates blocks N diffusion deeper in the bulk material. The presence of H neutralizes allowing components potentially acting as N traps. Finally, the incorporation of H to the ion beam slightly reduces the material hardness as compared the material implanted with a pure $N_2^+$ ion beam and influences the formed crystalline structure.

Acknowledgements

This work is part of the PhD thesis of LFZ. The authors are indebted to D. Ugarte for the SEM measure-ments. This work was partially sponsored by FAPESP, project #97/12069-0. LFZ and CAF are FAPESP fellows. FA is CNPq fellow. The electron microscopy work has been performed with JSM-5900LV microscope of the LME/LNLS, Campinas.